\documentclass{elsart}

\usepackage{amssymb}
\usepackage{amsmath}

\begin{document}

\begin{frontmatter}



\title{Relativistic two-body equation based on the extension of the $SL(2,C)$ group}

\author{D. A. Kulikov, }
\author{R. S. Tutik\corauthref{cor1}}
\ead{tutik@ff.dsu.dp.ua}
\author{and A. P. Yaroshenko}
\corauth[cor1]{corresponding author}

\address{Department of Physics,  Dniepropetrovsk National University, 13 Naukova str., Dniepropetrovsk 49050, Ukraine}

\begin{abstract}
A new approach to the two-body problem based on the extension of the $SL(2,C)$ group to the
$Sp(4,C)$ one is developed. The wave equation with various forms of including the interaction for the system of the spin-$1/2$ and spin-$0$ particles is constructed. For this system, it was found that the wave equation with a linear confinement potential involved in the non-minimal manner has an oscillator-like form and possesses the exact solution.
\end{abstract}

\end{frontmatter}

\newpage
\section{Introduction}
\label{part1}
Relativistic treatment of the two-body problem plays an important role in particle and nuclear physics. There have been developed various approaches to this problem because the Bethe-Salpeter equation \cite{bethe}, derived in the quantum field theory, has been exceedingly difficult to solve \cite{mainland}. Among them we have the relativistic quantum mechanics with constraints \cite{sazdjian86,crater87,crater06} and analogous equations obtained from the Bethe-Salpeter equation \cite{sazdjian85,sazdjian97,bijtebier}, the Barut method \cite{barut,klimek}, and, at last, we may directly postulate the Lorentz-invariant wave equations \cite{fushchych,nikitin}.

It has been proved that the Lorentz $SO(1,3)$ symmetry of the Minkowski space is locally isomorphic to the the symplectic $Sp(2,C)\equiv SL(2,C)$ symmetry \cite{penrose}. Recently, the symplectic extension \cite{pirogov} and also the general complex extension \cite{bogush} of the Lorentz group have been the object of detail investigations.

The goal of this work is to develop a new approach to the relativistic two-body problem, based on the extension of the $Sp(2,C)$ group to the $Sp(4,C)$ one. In this Letter we restrict ourselves only to the construction of the wave equation for the system composed of the spin-$1/2$ and spin-$0$ particles. This equation is suitable for the description of baryons as the quark-diquark systems.
Taking into account that the question about the Lorentz structure of the interquark confinement interaction is widely discussed in the literature \cite{olsson,simonov}, we give essential attention to introducing the confinement potential in various forms.
It has been found that the derived relativistic two-body equation with the linear confinement potential involved in a non-minimal manner possesses the exact solution.

The Letter is organized as follows. In Section 2 we outline the general procedure of the symplectic space-time extension. In Section 3 we apply above procedure for construction of the wave equation for the fermion-boson system without interaction. Section 4 is devoted to introducing the interaction in the minimal manner. In Section 5 we apply the developed technique of the symplectic space-time extension to introducing the linear confinement interaction in the non-minimal manner. For this interaction, we derive the wave equation for the system of spin-$1/2$ and spin-$0$ particles which has an exact solution. Finally, the results are summarized in the concluding section.

\section{Symplectic space-time extension}
\label{part2}
Let us recall that the symplectic group $Sp(2l,C)$, $l\geq 1$  is the group of $2l\times 2l$ matrices with complex elements and determinant equal to one \cite{weyl,hall}. $Sp(2l,C)$ matrices act on $2l$-component Weyl spinors and preserve an antisymmetric bilinear form which plays the role of "metrics" in the spinor space.

The homogeneous Lorentz group $SO(1,3)$ is covered by the $Sp(2,C)\equiv SL(2,C)$ group. As a consequence, there exists one-to-one correspondence between $Sp(2,C)$ Hermitian spin-tensors of second rank and four-vectors of the Minkowski space. It means that the space-time position of a relativistic particle can be parametrized by the $Sp(2,C)$ Hermitian spin-tensor. From this point of view, the description of a two-particle system requires a geometrical object composed of at least two $Sp(2,C)$ Hermitian spin-tensors.

In order to obtain such an object, we extend the $Sp(2,C)$ group to the $Sp(4,C)$ one. This minimal extension also preserves a non-degenerate
antisymmetric bilinear form $\eta_{\alpha\beta}=-\eta_{\beta\alpha}$ ($\alpha,\beta=1,2,3,4$) in the fundamental spinor space.
Then the $Sp(4,C)$ Weyl spinors $\varphi_{\alpha}$ with lower indices and their complex conjugatives $\bar{\varphi}_{\bar{\alpha}}=(\varphi_{\alpha})^*$ are related to spinors with upper indices by transformations
$\varphi_{\alpha}=\eta_{\alpha\beta}\varphi^{\beta}$ and
$\bar{\varphi}_{\bar{\alpha}}=\eta^*_{\bar{\alpha}\bar{\beta}}\bar{\varphi}^{\bar{\beta}}$.

As well as in the $Sp(2,C)$ group case, the $Sp(4,C)$ Hermitian spin-tensor,
$P_{\alpha\bar{\alpha}}$, corresponds to a real vector $P_M$. However, this vector now belongs to a Minkowski space $\mathsf{R^{16}}$ rather than to $\mathsf{R^{4}}$. Let us define this correspondence by
\begin{equation}\label{eq2}
P_{\alpha\bar{\alpha}}=\mu^M_{\alpha\bar{\alpha}}P_M, \qquad
P^M=\frac{1}{4}\tilde{\mu}^{M\bar{\alpha}\alpha}P_{\alpha\bar{\alpha}}
\end{equation}
where $\mu^M_{\alpha\bar{\alpha}}$ ($M=1\div 16$) are matrices of the basis in the space of $4 \times 4$
Hermitian matrices and tilde labels the transposed matrix with uppered spinorial indices. In what follows, the spinorial indices will be omitted when possible.

To clarify the relationship of the discussed vector space to the Minkowski space $\mathsf{R^{4}}$,
we represent 16 values of the vector index of $P_M$ through $4\times 4$ combinations of two indices, $M=(a,m)$, with both $a$ and $m$ running from $0$ to $3$. Then the matrices $\mu^M$
and $\tilde{\mu}^N$ can be chosen in the form
\begin{equation}\label{eq3}
\mu^M\equiv\mu^{(a,m)}=\Sigma^a\otimes\sigma^m,
\qquad \tilde{\mu}^N\equiv\tilde{\mu}^{(b,n)}=\tilde{\Sigma}^b\otimes\tilde{\sigma}^n,
\end{equation}
where explicit expressions for matrices $\Sigma^a$ and $\sigma^m$, written in terms of $2\times 2$ unit matrix $I$ and the Pauli matrices $\tau^i$, are
\begin{eqnarray}\label{eq4}
\Sigma^0=\tilde{\Sigma}^0=I, \quad \Sigma^1=\tilde{\Sigma}^1=\tau^1, \quad
\Sigma^2=-\tilde{\Sigma}^2=\tau^2, \quad \Sigma^3=\tilde{\Sigma}^3=\tau^3, \nonumber \\
\sigma^0=\tilde{\sigma}^0=I, \quad \sigma^1=-\tilde{\sigma}^1=\tau^1, \quad
\sigma^2=-\tilde{\sigma}^2=\tau^2, \quad \sigma^3=-\tilde{\sigma}^3=\tau^3.
\end{eqnarray}

Now the metrics of the discussed vector space
\begin{equation}\label{eq5}
g^{MN}=\frac{1}{4}\mu^M_{\alpha\bar{\alpha}}\tilde{\mu}^{N\bar{\alpha}\alpha}
\end{equation}
is reduced to the factorized form
\begin{equation}\label{eq6}
g^{MN}\equiv g^{(a,m)(b,n)}=\hat{h}^{ab}h^{mn}
\end{equation}
where $h^{mn}=diag(1,-1,-1,-1)$ is the usual Minkowski metrics and
$\hat{h}^{ab}=diag(1,1,-1,1)$ is caused by the group extension.

Since the factorization of the metrics means that the vector from $\mathsf{R^{16}}$ may be decomposed into four Minkowski four-vectors, it appears that the wave equation for the two-body problem may be constructed with the $Sp(4,C)$ momentum spin-tensor (\ref{eq2}).

\section{Construction of the wave equation for the fermion-boson system}
\label{part3}
In the present Letter we consider a system, consisted of one Dirac and one Klein-Gordon particle, with the total spin $1/2$. Its wave function can be represented by a Dirac spinor or, alternatively, by two Weyl spinors \cite{landau}. Due to the total spin of the system being equal to $1/2$, the wave equation must have the form
of the Dirac-like equation that in terms of $Sp(4,C)$ Weyl spinors reads
\begin{equation}\label{eq7}
P_{\alpha\bar{\alpha}}\bar{\chi}^{\bar{\alpha}}=m\varphi_{\alpha}, \qquad
\tilde{P}^{\bar{\alpha}\alpha}\varphi_{\alpha}=m\bar{\chi}^{\bar{\alpha}}
\end{equation}
where $P_{\alpha\bar{\alpha}}$ is the $Sp(4,C)$ momentum spin-tensor and $m$ is a mass parameter. Then, since each of spinors $\varphi_{\alpha}$ and $\bar{\chi}^{\bar{\alpha}}$ can be expanded into two $Sp(2,C)$ Weyl spinors, the proposed wave equation describes a doublet of spin-$1/2$ systems.

Further, for elucidating the two-particle interpretation of Eq.(\ref{eq7}), let us consider the structure of the $Sp(4,C)$ momentum spin-tensor. From Eqs.(\ref{eq2}) and (\ref{eq3}) it follows that
\begin{equation}\label{eq8}
P=\mu^{(a,m)}P_{(a,m)}=\Sigma^0\otimes\sigma^m w_m+\Sigma^1\otimes\sigma^m p_m+
\Sigma^2\otimes\sigma^m u_m+\Sigma^3\otimes\sigma^m q_m
\end{equation}
where $w_m$, $p_m$, $u_m$ and $q_m$ are the Minkowski four-momenta.

However, the description of the two-particle system requires only two four-momenta whereas the $Sp(4,C)$ momentum spin-tensor corresponds to four four-momenta, collected in a $16$-component vector.
Therefore we have to decrease the number of the independent components of
$w_m$, $p_m$, $u_m$ and $q_m$. For this, we begin with transforming our equation to the form of the Klein-Gordon equation. Eliminating $\bar{\chi}^{\bar{\alpha}}$ from Eq.(\ref{eq7}) and using Eq.(\ref{eq8}), we obtain
\begin{equation}\label{eq8a}
(P\tilde{P}-m^2)\varphi\equiv(w^2+p^2-u^2+q^2-m^2+\sum^{5}_{A=1}\gamma_A K^A)\varphi=0
\end{equation}
where $w^2=(w^0)^2-\mathbf{w}^2$, $p^2=(p^0)^2-\mathbf{p}^2$ etc, $\gamma_A$ are direct products of the Pauli matrices, and $K^A$ are quadratic forms with respect to the four-momenta.

Because the non-diagonal terms $\gamma_A K^A$ in Eq.(\ref{eq8a}) have no analog in the case of the ordinary Klein-Gordon equation, we put $\gamma_A K^A=0$ that yields
\begin{eqnarray}\label{eq8b}
&&wp+pw=0,\qquad wq+qw=0,\qquad up+pu=0,\qquad uq+qu=0, \nonumber \\
&&u^m w^n + w^n u^m - u^n w^m - w^m u^n-\epsilon^{mnkl}(p_k q_l + q_l p_k)=0
\end{eqnarray}
with $\epsilon^{mnkl}$ being the totally antisymmetric tensor ($\epsilon^{0123}=+1$).
The imposed conditions set ten components of $w_m$, $p_m$, $u_m$, $q_m$ to be
the independent ones.

Now we suggest that these four-momenta are expressed through the four-momenta, $p_{1m}$ and $p_{2m}$, of the constituent particles as follows
\begin{equation}\label{eq9}
w_m=\frac{1}{2}(p_{1m}+p_{2m}),\quad p_m=\frac{1}{2}(p_{1m}-p_{2m}),\quad u_m=0,\quad q_m=0.
\end{equation}
Then the only one condition from Eqs.(\ref{eq8b}) remains non-trivial that reads
\begin{equation}\label{eq9a}
wp+pw\equiv (p_1^2-p_2^2)/4=0.
\end{equation}
This equality implies that the total spinor wave function does not depend on the relative
time of the particles.

Further, if we use the wave equation (\ref{eq7}) and the condition (\ref{eq9a}), we may
derive the one-particle Dirac and Klein-Gordon equations for the constituents of our system. Indeed, with decomposing the spinor wave functions into the projections
\begin{equation}\label{eq10}
\varphi_{\pm}=\frac{1}{2}(1\pm\tau^1\otimes I)\varphi, \qquad
\bar{\chi}_{\pm}=\frac{1}{2}(1\pm\tau^1\otimes I)\bar{\chi}
\end{equation}
which are two-component $Sp(2,C)$ Weyl spinors as well, Eq.(\ref{eq7}) reduces to uncoupled sets of equations
\begin{equation}\label{eq11}
p_{1m}\sigma^m\bar{\chi}_{+}=m\varphi_{+}, \qquad
p_{1m}\tilde{\sigma}^m\varphi_{+}=m\bar{\chi}_{+}
\end{equation}
and
\begin{equation}\label{eq12}
p_{2m}\sigma^m\bar{\chi}_{-}=m\varphi_{-}, \qquad
p_{2m}\tilde{\sigma}^m\varphi_{-}=m\bar{\chi}_{-},
\end{equation}
which are the one-particle Dirac equations written in the Weyl spinor formalism \cite{landau}. Then the combination of Eqs.(\ref{eq9a}), (\ref{eq11}) and (\ref{eq12})
yields the one-particle  Klein-Gordon equations for all spinor projections $\varphi_{\pm}$, $\bar{\chi}_{\pm}$.
Hence, the total spinor wave function can be understood as the product of the spinor wave function of a fermion and the scalar wave function of a spin-$0$ particle.

Thus, we conclude that the wave equation (\ref{eq7}), based on the symplectic space-time extension and supplemented with the conditions (\ref{eq8b}), describes two systems composed of the spin-$1/2$ and spin-$0$ particles with equal masses $m_1=m_2=m$. These systems differ from each other only in permutation of the particles.

For describing the fermion-boson system with unequal masses, we should replace
the mass parameter in the right hand of the wave equation (\ref{eq7})
by $(m_1+m_2)/2+\tau^1\otimes I (m_1-m_2)/2$, that will be discussed elsewhere.

\section{Potential interaction involved in the minimal manner}
\label{part4}
Now we intend to include the potential interaction in the preceding results.
A common-used receipt consists in the replacement of the four-momenta of particles in the minimal manner by the generalized momenta, so that each particle is in an external potential of the other.

Upon inserting the substitution
\begin{equation}\label{eq17}
p_1^m\rightarrow\pi_1^m=p_1^m-A_1^m, \qquad p_2^m\rightarrow\pi_2^m=p_2^m-A_2^m
\end{equation}
into Eqs.(\ref{eq9}), the condition (\ref{eq9a}) transforms into $(\pi_1^2-\pi_2^2)/4=0$
that must be compatible with the wave equation (\ref{eq7}). It means that the operator
$\pi_1^2-\pi_2^2$ must commute with the operators acting on the wave functions in Eq(\ref{eq7}). The compatibility is achieved by taking
\begin{equation}\label{eq18}
\pi_1^2-\pi_2^2=p_1^2-p_2^2,
\end{equation}
that obviously restricts the shape of the potentials in $\pi_1^m$ and $\pi_2^m$.
Furthermore, the potentials have to depend on the relative coordinate $x=x_1-x_2$ only through its transverse with respect to the total momentum part.

With the generalized momenta, obeying Eq.(\ref{eq18}), we can introduce both the potential  interaction described by the time-component of the Lorentz vector and the confinement one, proposed in the two-particle relativistic quantum mechanics with constraints \cite{sazdjian86,crater87}.

It should be mentioned that the confinement interaction may be involved into the wave equation (\ref{eq7}) by means of the scalar potenial ($S$) through the substitution $m\rightarrow m+S$, too.

However, the discussed symplectic space-time extension enables to introduce not only above mentioned types of the potential interactions but also the ones involved in the non-minimal manner as in the Dirac oscillator model \cite{moshinsky,kukulin}. 

\section{Interaction involved in the non-minimal manner. A solvable example}
\label{part5}
As an application of the discussed method, let us consider the two-body problem for the system of the spin-$1/2$ and spin-$0$ particles interacting through the linearly rising confinement potential included in the wave equation in the non-minimal manner, i.e. in another way than the minimal replacement of the four-momenta of particles by the generalized momenta (\ref{eq17}). This problem, apart from the academic interest as possessing an exact solution, has the practical importance, being a simple quark-diquark model for baryons.

In the construction of the wave equation for such a system, instead of the minimal replacement for the momenta, the interaction will be included in the subsidiary momenta $q_m$ and $u_m$ which are zeroth in the free particles case. To satisfy the conditions (\ref{eq8b}), we leave the expressions (\ref{eq9}) for the total and relative momenta $w_m$ and $p_m$ without changes but set for the subsidiary ones
\begin{equation}\label{eq19}
\quad q_m=\lambda x_{\bot m},\qquad u_m=\frac{1}{2w^2}\epsilon_{mnkl}w^n(p^k q^l + q^l p^k)
\end{equation}
where  $x=x_1-x_2$, which transverse with respect to the total momentum part is defined as $x_{\bot}^m=(h^{mn}-w^m w^n/w^2)x_n$, and $\lambda$ is a constant. Thus, we involve the linear confinement interaction in the non-minimal manner.

Supposing the total momentum $w_m$ to be conserved, conditions (\ref{eq8b}) for the momenta $w_m$, $p_m$, $u_m$, $q_m$ reduce to Eq.(\ref{eq9a}) that is compatible with the wave equation (\ref{eq7}).
The wave equation with such interaction can be solved exactly. To find a solution, we use the center-of-mass frame in which $\mathbf{w}=0$. Then $2w^0=E$ is the
total energy and $\mathbf{x}_{\bot}=\mathbf{x}\equiv \mathbf{x}_1-\mathbf{x}_2$ is the interparticle separation. From Eq.(\ref{eq9a}) it follows that the zeroth component of the relative four-momentum vanishes, $p^0=0$, and
the final expression for the momentum spin-tensor (\ref{eq8}) becomes
\begin{equation}\label{eq20}
P=\frac{E}{2}I\otimes I-\tau^1\otimes\boldsymbol\tau\mathbf{p}
-\lambda\tau^3\otimes\boldsymbol\tau\mathbf{x}-\frac{2\lambda}{E}\tau^2\otimes
\boldsymbol\tau\mathbf{l}
\end{equation}
where $\mathbf{l}=\mathbf{x}\times\mathbf{p}$ is the orbital angular momentum of the relative motion.

It is easy to verify that the total angular momentum $\mathbf{j}=\mathbf{l}+\boldsymbol\tau/2$ and the spin-orbit coupling operator $K=\tau^2\otimes(\boldsymbol\tau\mathbf{l}+I)$ are conserved. Since the latter can be replaced by its eigenvalues $\kappa=\pm(j+1/2)$, Eq.(\ref{eq7}) transforms into the oscillator-type equation for the function $\psi=\varphi+\bar{\chi}$
\begin{equation}\label{eq21}
[(\frac{E}{2}-\frac{2\lambda\kappa}{E})^2-(m-\frac{2\lambda}{E}\tau^2\otimes I)^2]\psi=
[\mathbf{p}^2+\lambda^2\mathbf{x}^2-\lambda(2\kappa+\tau^2\otimes I)]\psi.
\end{equation}

The exact physical solution to this equation written in the form of the Regge trajectories for the bound states of the fermion-boson system is
\begin{eqnarray}\label{eq22}
&&j=\frac{E^2}{4|\lambda|}-\frac{1}{2}-\sqrt{\frac{E^2}{4|\lambda|}(4n+1-\frac{\lambda}{|\lambda|})+\left(\frac{mE}{2\lambda}-1\right)^2},
\quad \kappa=+(j+\frac{1}{2}), \nonumber \\
&&j=\frac{E^2}{4|\lambda|}-\frac{1}{2}-\sqrt{\frac{E^2}{4|\lambda|}(4n+3-\frac{\lambda}{|\lambda|})+\left(\frac{mE}{2\lambda}-1\right)^2},
\quad \kappa=-(j+\frac{1}{2})
\end{eqnarray}
where $n=0,1,2,...$ is the radial quantum number, and $\lambda=\pm|\lambda|$ corresponds to the contributions of the spin-orbit terms of opposite signs.

Since the derived Regge trajectories are nearly linear in the squared energy and have
the same slope in asymptotics, we conclude that the obtained equation is reasonable to be considered as a simplified quark-diquark model for baryons.  The description of the baryon spectra within this approach will be published elsewhere.

\section{Summary}
\label{part6}
A new approach to the two-body problem based on the extension of the $SL(2,C)$ group to the
$Sp(4,C)$ one has been developed. This approach permits us to construct the relativistic wave equation for the system consisted of spin-$1/2$ and spin-$0$ particles, involving the various forms of interaction. For this system, the exact solution of the wave equation with a linear confinement potential introduced in the non-minimal manner is found. Obtained Regge trajectories have the same behavior as the behavior of the baryonic Regge trajectories that permits us to consider this wave equation as a simplified quark-diquark model for baryons. The description of the baryon spectra within the more realistic quark-diquark model with unequal masses of constituents and including the confinement interaction in various ways will be published elsewhere.

\section{Acknowledgements}
This research was supported by a grant N 0106U000782 from the Ministry of Education and Science of Ukraine which is gratefully acknowledged. The authors thank the anonymous referee for his comments aimed at improving the Letter.



\end{document}